\documentclass[12pt,reqno]{amsart}
\advance\voffset by -2.0cm \advance \hoffset by -1.5cm
\usepackage{hyperref}

\usepackage{cite}
\usepackage{amsmath}
\usepackage{amssymb}
\usepackage{amsthm}
\usepackage{graphicx,color}
\usepackage{verbatim}

\usepackage{mathtools}

\usepackage{bbm}

\newcommand{\RR}{\mathbb{R}} 
\newcommand{\QQ}{\mathbb{Q}}

\newcommand{\ZZ}{\mathbb{Z}} 
\newcommand{\NN}{\mathbb{N}} 
 
\newcommand{\MM}{\mathbb{M}}

\newcommand{\C}{\mathcal{C}}

\newcommand{\N}{\mathcal{N}}

\newcommand{\T}{\mathcal{T}}

\usepackage{xcolor}

\newtheorem{definition}{Definition}

\DeclareMathOperator{\Aut}{Aut}

\numberwithin{equation}{section}

\def\be{\begin{equation}}
\def\ee{\end{equation}}
\newcommand{\bea}{\begin{eqnarray}}
\newcommand{\eea}{\end{eqnarray}}

\allowdisplaybreaks[3]
\begin{document}
\title[Dualities in CHL-Models]{Dualities in CHL-Models}
%\author{Natalie Paquette, Daniel Persson, Roberto Volpato}

\author{Daniel Persson}
\address{Department of  Mathematical Sciences, Chalmers University of Technology,
412 96, Gothenburg, Sweden}
\email{daniel.persson@chalmers.se}

\author{Roberto Volpato}
\address{Dipartimento di Fisica e Astronomia `Galileo Galilei', Universit\`a di Padova \& INFN, sez. di Padova, Via Marzolo 8, 35131, Padova, Italy}
%\newline
%\indent Theory  Group,  SLAC  National  Accelerator  Laboratory, Menlo  Park,
%CA 94025, USA  \newline
%\indent Stanford  Institute  for  Theoretical  Physics,  Department  of  Physics,  Stanford 
% University, Stanford, CA 94305, USA}
\email{volpato@pd.infn.it}

\begin{abstract} 
We define a very general class of CHL-models associated with any string theory  (bosonic or supersymmetric) compactified on an internal CFT $\mathcal{C}\times T^d$. We take the orbifold by a pair $(g,\delta)$, where $g$ is a (possibly non-geometric) symmetry of $\mathcal{C}$ and $\delta$ is a translation along $T^d$.  We analyze the T-dualities of these models and show that in general they contain Atkin-Lehner type symmetries. This generalizes our previous work on $\mathcal{N}=4$ CHL-models based on heterotic string theory on $T^6$ or type II on $K3\times T^2$, as well as the `monstrous' CHL-models based on a compactification of heterotic string theory on the Frenkel-Lepowsky-Meurman CFT $V^{\natural}$.
\end{abstract}
\maketitle

%\vspace{2cm}
%\newpage
\tableofcontents

\section{Introduction}
\noindent String theories with 16 supersymmetries have served as an important arena for probing the quantum structure of BPS black holes. In the simplest instance of heterotic string theory on $T^6$,  Dijkgraaf, Verlinde and Verlinde proposed a remarkable formula for the exact microscopic degeneracies of 1/4-BPS states, in terms of the Fourier coefficients of (the inverse of) a certain Siegel modular form $\Phi_{10}$, the so called Igusa cusp form of weight 10. This formula was later proven by explicit counting of states in the dual type IIA/$K3\times T^2$-picture \cite{Shih:2005uc}, and has also been verified by explicit calculations in supergravity using localization \cite{Dabholkar:2011ec,Dabholkar:2010uh,Dabholkar:2014ema,Murthy:2015yfa,Murthy:2015zzy}. The contributions from single-centered states, i.e. black holes, can be isolated via a  canonical decomposition of the Fourier coefficients of $1/\Phi_{10}$, and are controlled by a certain mock modular form \cite{Dabholkar:2012nd}.

A larger class of $\mathcal{N}=4$ string theories were constructed by Chaudhuri, Hockney and Lykken \cite{Chaudhuri:1995fk}, now generally referred to as CHL-models. These models are obtained by taking orbifolds of heterotic string theory on $T^6$ (or, dually, of type IIA on $K3\times T^2$) by geometric symmetries of the target space. It has been shown that the counting of 1/4 BPS-states in CHL-models can be accounted for by the Fourier coefficients of a class of Siegel modular forms $\Phi_N$, parametrized by the order $N$ or the orbifold symmetry \cite{David:2006ji,Sen:2009md,Sen:2010ts,Govindarajan:2010fu,Govindarajan:2011em}. 

In the context of (generalised) Mathieu moonshine \cite{Cheng:2010pq,Gaberdiel:2012gf,Gaberdiel:2013nya,Persson:2013xpa}, a large class of Siegel modular forms $\Phi_{g,h}$ were constructed, parametrized by commuting pairs of elements $g,h\in M_{24}$. When restricting to elements $(1,h)$ with $h$ an order $N$ geometric symmetry of $K3\times T^2$ these Siegel modular forms restrict to the partition functions $\Phi_N$ of CHL-models. This suggested that there should exist a more general class of CHL-models associated with non-geometric symmetries $g$, whose BPS-states are counted by the Siegel modular forms $\Phi_{g,h}$. These non-geometric CHL-models were constructed in our previous work \cite{Persson:2015jka} using the following prescription. Consider type IIA string theory on $K3\times T^2$ and decompose the torus according to $T^2=S^{1}\times \tilde{S}^1$. We then orbifold this theory by a pair $(g,\delta)$, where $\delta$ is an $n$:th order translation along one of the circles, and $g$ is an order $N$ symmetry of the internal $\mathcal{N}=(4,4)$ superconformal sigma model on $K3$. Any such symmetry necessarily belongs to the Conway group $Co_0$ \cite{K3symm}. We showed that the T-duality group for these theories contains the Fricke involution $T\to -1/(NT)$, where $T$ is K\"ahler modulus of $T^2$. This was quite surprising since the Fricke involution is not contained in the the original $SL(2,\mathbb{Z})$ T-duality group of the parent (unorbifolded) string theory on $K3\times T^2$. Recently Fricke dualities were analyzed in the context of higher-derivative couplings in heterotic CHL-models \cite{Bossard:2017wum}. This Fricke symmetry also turns out to have interesting consequences for the lattice of electric-magnetic charges $L=L_e\oplus L_m$, namely that they should be $N${\it-modular}:
\begin{equation}
L_e\cong L_e^{*}(1/N),\qquad L_m\cong L_m^{*}(N),
\end{equation}
where $L^{*}$ denotes the dual lattice and $L(N)$ is a rescaling of the quadratic form by $N$.

This construction of CHL-models was later used in \cite{Paquette:2016xoo} to give a new supersymmetric interpretation of monstrous moonshine. Here, one considers heterotic string theory on $V^\natural\times \bar{V}^{s\natural}\times T^2$, where $V^\natural$  is the Frenkel-Lepowsky-Meurman monster CFT \cite{FLM}, and $\bar{V}^{s\natural}$ is Duncan's superconformal field theory \cite{duncan2007super} which yields a moonshine module for the Conway group $Co_0$. We then orbifold by a pair $(g,\delta)$, where $g$ an element of the monster $\mathbb{M}$ and $\delta$ a translation along one of the circles of $T^2$. In this context, the somewhat mysterious Fricke involutions of monstrous moonshine appear naturally as spacetime T-dualities in the monstrous CHL-model.

In view of the examples discussed above it is interesting to ask in what context the CHL-construction applies and what is the associated T-duality group. The main point of the present paper is to discuss this question by defining a very general class of CHL-model and analyze some of its properties. We begin in section \ref{sec_defCHL} to define these CHL-models in full generality for any string theory (type I, type II, heterotic). We then analyze their spectrum and classify the structure of the associated duality groups. In section \ref{sec:TS} we consider Atkin-Lehner type dualities and show that for self-dual models they require $N$-modularity of the electric-magnetic charge lattices, generalizing our previous work \cite{Persson:2015jka}. In the concluding section \ref{sec_disc} we offer some suggestions for interesting future research.

 \section{General CHL models}
 \label{sec_defCHL}

\noindent In this section we describe some general features of CHL models. CHL models were first considered by Chaudhuri, Hockney and Lykken in the context of superstring compactifications with $16$ spacetime supersymmetries \cite{Chaudhuri:1995fk}. However, their general structure applies to many other cases, so it is useful to define them more generally.

\subsection{Definition}
\label{sec:CHLdef}
The starting point in the construction of a CHL model is a compactification of some string theory $\mathcal{S}$ on $\C\times T^d$, where $\C$ is a  CFT and $T^d$ is a $d$-dimensional torus (we will mostly be interested in $d=1,2$). In most examples, $\C$ will be a non-linear sigma on some manifold $X$; in this case it will be denoted by $\C^X$. Note that we allow for any string theory, so $\mathcal{S}$ can denote a bosonic, heterotic, type I or type II string theory. The CFT $\C\times T^d$ is assumed to have the right central charges and number of world-sheet supersymmetries to provide a consistent compactification of the string theory $\mathcal{S}$. Denote this string compactification by $\mathcal{S}[\C\times T^d]$. Let $\delta$ be a translation by $1/N$, for some $N>0$, along a circle $S^1\subset T^d$, and let $g$ be an order $N$ finite automorphism of the CFT $\mathcal{C}$ which commutes with all spacetime supersymmetries and exists at generic points in the moduli space of $\mathcal{C}$. We then have:

\begin{definition}
We define the CHL-model associated with the string theory $\mathcal{S}[\C\times T^d]$ to be the orbifold
\begin{equation}
\mathbb{CHL}_{g,\delta}(\mathcal{S}, \C, T^d):=\mathcal{S}[\C\times T^d]\big/\left<(g, \delta)\right>,
\end{equation}
where we orbifold simultaneously $g$ on $\C$ and by $\delta$ on $T^d$.
\end{definition}

In general, the orbifold of $\mathcal{C}$ will not satisfy the level-matching condition. The CHL-model is still well defined, however, since the failure of level-matching can be compensated by the shift of $\delta$. The general condition for level-matching to be satisfied is that in the $g$-twisted sectors one has:
\begin{equation}
L_0-\bar L_0\in \frac{1}{N(g)}\ZZ,
\end{equation}
where $N(g)$ is the order of $g$.

In principle we could have allowed for $g$ and $\delta$ to have different orders. However, without loss of generality one can always assume that $g$ and $\delta$ have the same order $N$. Indeed, suppose that $\delta$ has order $M$ and $g$ has order $N$. One then has 
\begin{equation}
\mathbb{CHL}_{g,\delta}(\mathcal{S}, \C, T^d)= \mathbb{CHL}_{g',\delta'}(\mathcal{S}, \hat{\C}, \hat{T}^d), 
\end{equation}
where $g',\delta'$ both have order $\gcd(M,N)$ and we defined:
\begin{equation}
\hat{\C}=\C\big/\left<g^{M/\gcd(M,N)}\right>, \qquad \hat{T}^d=T^d\big/\left<\delta^{N/gcd(M,N)}\right>.
\end{equation}
This implies that one can always reduce to the case where $\delta$ and $g$ have the same order.

%taking the CHL orbifold of $\C\times T^d$ by $(\delta,g)$, with $\delta$ and $g$ of orders $N$ and $M$, respectively, is the same thing as taking the CHL orbifold of $\C'\times {T'}^d$ by $(\delta',g')$ where both symmetries have order $\gcd(N,M)$. Here $\C'$ and ${T'}^d$ are the orbifolds of $\C$ and $T^d$ by $g^{\frac{M}{\gcd(N,M)}}$ and $\delta^{\frac{N}{\gcd(N,M)}}$, respectively. 

To illustrate the general construction let us now consider some examples.

\vspace{.3cm}
\noindent {\bf Example 1.}
Consider type IIA superstring theory compactified on $K3\times T^2$. Then $\C^X=\C^{K3}$ is an $\mathcal{N}=(4,4)$ non-linear sigma model on K3. We take the orbifold by $(\delta,g)$, where $\delta$ is a shift by $1/N$ of a period along a circle $S^1\subset T^2$ and $g$ is a symmetry  of $\C^{K3}$ of order $N$ preserving the $\N=(4,4)$ superconformal algebra and the half-integral spectral flows. With these properties, the resulting CHL model $\mathbb{CHL}_{g,\delta}(\text{IIA}, \C^{K3}, T^2)$ is a four dimensional $\N=4$ theory, where the number of gauge multiplets depends on the symmetry $g$. By T-duality on the non-orbifolded circle in $T^2=S^1\times \tilde{S}^1$ this construction also yields an equivalent CHL-model based on the type IIB superstring. 
\vspace{.4cm}

\noindent {\bf Example 2.}
Another example can be readily obtained by heterotic-type II string duality. Indeed,  type IIA string theory on $K3\times T^2$ is equivalent to heterotic string theory on $T^4\times T^2$ and, by the adiabatic argument of Vafa-Witten \cite{VafaWitten95}, this duality commutes with taking the CHL orbifold. This means that  the CHL models described in the previous point are dual to a CHL model based on the symmetry $(\delta,g)$, where $g$ is now a symmetry of the heterotic strings on $T^4$ commuting with the right-moving $N=4$ supersymmetry:
\begin{equation}
\mathbb{CHL}_{g,\delta}(\text{Het}, \C^{T^4}, T^2)\longleftrightarrow \mathbb{CHL}_{g,\delta}(\text{IIA}, \C^{K3}, T^2)\longleftrightarrow \mathbb{CHL}_{g,\delta}(\text{IIB}, \C^{K3}, T^2)
\end{equation}

\vspace{.4cm}

\noindent {\bf Example 3.}
Let us now consider a rather different example. Let $\mathcal{S}$ be the heterotic string and compactiy this on $\C\times S^1$, where $\C=V^\natural\times \bar V^{s\natural}$ is the product of the Frenkel-Lepowski-Meurman (FLM) Monster module $V^\natural$ of central charge $c=24$, with automorphism group the Monster group $\MM$, and $\bar V^{s\natural}$ is a (anti-holomorphic) super-VOA of central charge $12$ with  automorphism group the Conway group $Co_0$. Consider now the pair $(g, \delta)$ where $g\in \MM$ and $\delta$ is a shift on $S^1$. In a similar vein as before, one can then define CHL models $\mathbb{CHL}_{g,\delta}(\text{Het}, V^\natural\times \bar V^{s\natural}, S^1)$. We stress that here $g$ only acts on $V^\natural$, and  hence the super-VOA $V^{s\natural}$ is merely a spectator in the orbifold process, required to make sure that the resulting CHL-model is supersymmetric. This class of CHL-models were introduced and studied in \cite{Paquette:2016xoo}.\footnote{In \cite{Paquette:2016xoo} a $0$-dimensional Euclidean version of this orbifold (i.e., with starting point $\C\times T^2$) was also considered.} They  were in particular used to provide a physical understanding of the Hauptmodul properties of Monstrous moonshine.

\vspace{.4cm}

\subsection{Spectrum}

Let us now analyze some more properties of the general CHL-models $\mathbb{CHL}_{g,\delta}(\mathcal{S}, \C, T^d)$. We assume first that the orbifold of $\C$ by $g$ satisfies the level matching condition, i.e. the $g$-twisted sectors satisfy $L_0-\bar L_0\in \frac{1}{N(g)}\ZZ$. The general case is slightly more complicated and will be considered later on. Let us also assume that $g$ and $\delta$ have the same order $N$. 

%This is not much of a restriction. Indeed, taking the CHL orbifold of $\C\times T^d$ by $(\delta,g)$, with $\delta$ and $g$ of orders $N$ and $M$, respectively, is the same thing as taking the CHL orbifold of $\C'\times {T'}^d$ by $(\delta',g')$ where both symmetries have order $\gcd(N,M)$. Here $\C'$ and ${T'}^d$ are the orbifolds of $\C$ and $T^d$ by $g^{\frac{M}{\gcd(N,M)}}$ and $\delta^{\frac{N}{\gcd(N,M)}}$, respectively. So, one can always reduce to the case where $\delta$ and $g$ have the same order.

For any CFT $\C$, we denote by $\C_{g^r}$ the $g^r$-twisted sector of $\C$, $r\in \ZZ/N\ZZ$, and define
\be \C_{r,s}:=\{\psi\in \C_{g^r}\mid g(\psi)=e^{-\frac{2\pi i s}{N}}\psi\}
\ee the eigenspace with eigenvalue $g=e^{-\frac{2\pi i s}{N}}$ in the $g^r$-twisted sector.

%Notice that states in $\C^{K3}_{r,s}$ have $g$-eigenvalue $e^{-\frac{2\pi i s}{N}}$ while states in sign $\C^{T^2}_{r,s}$ have $\delta$-eigenvalue $e^{\frac{2\pi i s}{N}}$, so that states in $\C^{K3}_{r,s}\otimes \C^{T^2}_{r,s}$ are $(\delta,g)$-invariant.
We will now consider the orbifold of $\C$ by a symmetry $g$ and of the torus model $\C^{T^d}$ by the translation $\delta$. The corresponding eigenspace decompositions of the associated $g^r$- and $\delta^r$-twisted sectors are given by: 
\begin{eqnarray} 
\C_{r,s}&=&\{\psi \in \C_{g^r}\mid g(\psi)=e^{-\frac{2\pi i s}{N}}\psi\},
\nonumber \\
\C^{T^d}_{r,s}&:=&\{\psi\in \C^{T^d}_{\delta^r}\mid \delta(\psi)=e^{\frac{2\pi i s}{N}}\psi\}\ .
\end{eqnarray}
Note that since the eigenvalues are equal up a sign, the states in the product 
\begin{equation}
\C_{r,s}\otimes \C^{T^d}_{r,s}
\end{equation}
are $(g, \delta)$-invariant. 

The spectrum of the associated CHL model can now be described as follows
\be\label{decompo} 
\mathbb{CHL}_{g,\delta}(\mathcal{S}, \C, T^d)=\bigoplus_{(r,s)\in \ZZ/N\ZZ\times \ZZ/N\ZZ} \C_{r,s}\otimes \C^{T^d}_{r,s}\ .
\ee 
Physically, $\C_{r,s}\otimes \C^{T^d}_{r,s}$ is the $\delta$-eigenspace with eigenvalue $e^{\frac{2\pi is}{N}}$ (or, equivalently, the $g$-eigenspace with eigenvalue $e^{\frac{-2\pi is}{N}}$ twisted sector), corresponding to the $r$-twisted sector in the CHL model.

\medskip

It is fruitful to give an alternative description of the spectrum from the point of view of the underlying lattices. The Narain lattice $L$ of winding and momenta along $T^d$ in the original $\C\times T^d$ compactification is given by
\be L\cong \Gamma^{d,d}\ ,
\ee where $\Gamma^{d,d}$ is the unique (up to isomorphisms) even unimodular lattice of signature $(d,d)$. It can be described as $\Gamma^{d,d}=(\Gamma^{1,1})^{\oplus d}\cong \ZZ^{2d}$, where $\Gamma^{1,1}\cong \ZZ\oplus \ZZ$ with quadratic form $\left(\begin{smallmatrix}
0 & 1\\ 1 & 0
\end{smallmatrix}\right)$.

Notice that $\delta$ can be defined as a null vector in $\frac{1}{N}L^\vee$; the corresponding shift is just the symmetry that multiplies a state with charges $\gamma\in L$ by $e^{2\pi i (\delta,\gamma)}$.\footnote{Here, $L^\vee$ is the dual of $L$. Of course, being $L$ unimodular, one has $L\cong L^\vee$. However, we prefer to keep this notation because this might be useful for further generalizations. } The symmetry defined by $\delta$ acts trivially if and only if $\delta\in L^\vee\subseteq \frac{1}{N}L^\vee$, so the CHL model only depends on the class $[\delta]$ in the quotient $\frac{1}{N}L^\vee/L^\vee$. All such classes $[\delta]$ are related to each other by automorphisms of the lattice (i.e., T-dualities of the model), so we can simply take $\delta$ to be the vector ${}^t(\frac{1}{N},0,\ldots,0)\in \frac{1}{N}L\cong (\frac{1}{N}\ZZ)^{2d}$. Roughly speaking, this corresponds to taking the shift $\delta$ along the circle $S^1\subseteq T^d$ corresponding to the first $\Gamma^{1,1}$ summand in $\Gamma^{d,d}=\Gamma^{1,1}\oplus \Gamma^{d-1,d-1}$. The sublattice of $\delta$-invariant vectors in $L$ is given by
\be L^{(\delta)}_{0,0}:=\{v\in L\mid (v,\delta)\in \ZZ\}=\ZZ\oplus N\ZZ\oplus \Gamma^{d-1,d-1}\ ,
\ee and (as the notation suggests) is the lattice of winding-momenta in the untwisted $\delta$-invariant sector $\C_{0,0}\otimes \C^{T^d}_{0,0}$ of the CHL model. The full lattice $L^{(\delta)}$ of winding-momenta of the CHL model is given by the dual of $L^{(\delta)}_{0,0}$. It is clear that such a dual lattice contains $\delta$; in fact,  $L^{(\delta)}$ is generated by $L$ and $\delta$, so that
\be L^{(\delta)}\equiv (L^{(\delta)}_{0,0})^\vee =\frac{1}{N}\ZZ\oplus \ZZ\oplus \Gamma^{d-1,d-1}\ .
\ee 
The lattice $L^{(\delta)}$ has a natural decomposition as the union of $L^{(\delta)}/L^{(\delta)}_{0,0}$ cosets
\be L^{(\delta)}=\bigcup_{r,s\in \ZZ/N\ZZ} L^{(\delta)}_{r,s}\ ,
\ee where
\be L^{(\delta)}_{r,s}:=\{r\delta+v\mid v\in L,\ (\delta,v)\equiv s\mod N\} \ .
\ee More precisely, each $L^{(\delta)}_{r,s}$ is the set of winding-momenta of the sector $\C_{r,s}\otimes \C^{T^d}_{r,s}$. 

We can now rewrite the spectral decomposition of the CHL model as a direct sum of sectors labeled by $L^{(\delta)}/L^{(\delta)}_{0,0}\cong \ZZ/N\ZZ\times \ZZ/N\ZZ$, i.e.
\be \mathbb{CHL}_{g,\delta}(\mathcal{S}, \C, T^d)=\bigoplus_{(r,s)\in L^{(\delta)}/L^{(\delta)}_{0,0}} \C_{r,s}\otimes \C^{T^d}_{r,s}\ .
\ee  The group $L^{(\delta)}/L^{(\delta)}_{0,0}$ has a quadratic form $q:L^{(\delta)}/L^{(\delta)}_{0,0}\to \QQ/2\ZZ$ induced by the quadratic form on $L^{(N)}$ and defined by $q([v])=(v,v)$ for some representative $v$ of $[v]\in L^{(\delta)}/L^{(\delta)}_{0,0}$. Explicitly, $q(r,s)=\frac{2rs}{N}$, where $(r,s)\in \ZZ/N\ZZ\times \ZZ/N\ZZ$.

\subsection{T-dualities of CHL models}\label{s:CHLTdual}

%We want to study the group of T-dualities of the CHL model. As a warm up, we consider the simpler case where $V=0$, so that we don't care about the charge $l$. This means that the interesting lattice of charges is 
%\be\begin{pmatrix}
% m_\parallel & m_t  & w_t & w_\parallel
%\end{pmatrix}\in M^{(N)}:=\ZZ\oplus\ZZ\oplus\ZZ\oplus\frac{1}{N}\ZZ
%\ee with quadratic form
%\be \begin{pmatrix}
%0 & 0 &  0& 1\\
%0 & 0 &  1& 0\\
%0 & 1 &  0& 0\\
%1 & 0 &  0& 0
%\end{pmatrix}
%\ee of signature $(2,2)$. This is the dual $M^{(N)}=(M^{(N)}_{0,0})^\vee$ of the sublattice of winding-momenta of the untwisted $g$-invariant subsector
%\be M^{(N)}_{0,0}\cong N\ZZ\oplus\ZZ\oplus\ZZ\oplus\ZZ\ .
%\ee Then the spectrum of the theory is given by a sum over the cosets labeled by $M^{(N)}/M^{(N)}_{0,0}\cong \ZZ_N\times \ZZ_N$.
%
%\bigskip

The T-duality group of the CHL model is the group of automorphisms of the lattice $L^{(\delta)}$. This corresponds to the subgroup of transformations in $ O(d,d,\RR)$  that preserve the lattice $L^{(\delta)}$. We are mostly interested in the case $d=2$, though we will keep the discussion general.

Notice that $O(2,2,\RR)$ has four connected components, and the component $SO^+(2,2,\RR)$ connected to the identity is isomorphic to 
\be SO^+(2,2,\RR)\cong (SL(2,\RR)\times SL(2,\RR))/(-1,-1)\ .\ee   The action of $(\left(\begin{smallmatrix}
a & b\\
c & d
\end{smallmatrix}\right),\left(\begin{smallmatrix}
a' & b'\\
c' & d'
\end{smallmatrix}\right)) \in SL(2,\RR)\times SL(2,\RR)$ on the vectors $(w_\parallel,m_\parallel,w_t,m_t)\in \RR^{2,2}$ is given by
\be \begin{pmatrix}
w_t & w_\parallel\\
-m_\parallel & m_t
\end{pmatrix}\mapsto \begin{pmatrix}
d & -b\\
-c & a
\end{pmatrix}\begin{pmatrix}
w_t & w_\parallel\\
-m_\parallel & m_t
\end{pmatrix}\begin{pmatrix}
a' & b'\\
c' & d'
\end{pmatrix}\ .
\ee

Rather than studying the full T-duality group it will be convenient to distinguish between different automorphism subgroups which we list below. 

\vspace{.3cm}

\subsubsection{Non-self-dualities}
\noindent Consider the   of automorphisms in $\Aut(L^{(\delta)})$ that can (possibly) act non-trivially on $L^{(\delta)}/L^{(\delta)}_{0,0}$. For $d=2$ this corresponds to those $\gamma\in \Aut(L^{(\delta)})$ such that 
$\gamma (r,s)\neq (r,s)\in \ZZ/N\ZZ\times \ZZ/N\ZZ$.  More precisely, $\Aut(L^{(\delta)})$ acts on  $\ZZ/N\ZZ\times \ZZ/N\ZZ$ by automorphisms that preserve the quadratic form $q(r,s)=\frac{2rs}{N}\mod 2\ZZ$. These maps permute the various sectors $\C^{T^2}_{r,s}$.  Since each such sector is tensored with the CFT factor $\C_{r,s}$ and, in general, $\C_{\gamma (r,s)}\neq \C_{r,s}$,  the elements of $\Aut(L^{(\delta)})$ are generically \emph{not} self-dualities of the CHL model. In fact, one can always find a model $\C'$ and a symmetry $g'$ such that 
$\C_{\gamma (r,s)}=\C'_{r,s}$. In particular, $\C'$ is the CFT with spectrum $\oplus_{s\in \ZZ/N\ZZ} \C_{\gamma(0,s)}$ and $g'$ is the symmetry acting by multiplication by $e^{-\frac{2\pi i s}{N}}$ on $\C_{\gamma(0,s)}\subseteq \C'$.
This means that, generically, $\gamma$ is a duality between the  CHL model based on $(\C,g)$ and the CHL model based on $(\C',g')$. For $d=2$, the group $\Aut(L^{(\delta)})$ has been calculated to be
\begin{multline} \Aut(L^{(\delta)})=\{\bigl(\frac{1}{\sqrt{e}}\begin{pmatrix}
ae & b\\ cN & de
\end{pmatrix}, \frac{1}{\sqrt{e}}\begin{pmatrix}
a'e & b'\\ c'N & d'e
\end{pmatrix}\bigr)\in SL(2,\RR)\times SL(2,\RR)\\ a,b,c,d,a',b',c',d'\in \ZZ,\ e\in \ZZ_{>0},\ e||N\}\ .
\end{multline}
This is generated by $\Gamma_0(N)\times \Gamma_0(N)$ and by $(W_e,W_e)$ for all exact divisors $e||N$, i.e. those $e\in \NN$ such that $e|N$ and $e\nmid \frac{N}{e}$. Here, $W_e$ is the Atkin-Lehner involution, an element $W_e\in SL(2,\mathbb{R})$ given by 
\begin{equation}
W_e=\frac{1}{\sqrt{e}}\begin{pmatrix} ae & b \\ Nc & de\end{pmatrix},
\end{equation}
such that 
\begin{equation}
a, b, c, d, e, N\in \mathbb{Z}, \qquad ade^2-Nbc=e, \qquad e||N.
\end{equation}
Any such element satisfies $W_e^2\in \Gamma_0(N)$ and is such that $W_e\Gamma_0(N)W_e^{-1}=\Gamma_0(N)$. In the special case $e=N$, $W_N$ can be taken of the form
\be W_N=\begin{pmatrix} 0 & -1/\sqrt{N} \\ \sqrt{N} & 0\end{pmatrix}
\ee which is known as Fricke involution. For Atkin-Lehner involutions, the model $\C'$ is the orbifold of $\C$ by $\langle g^{N/e}\rangle$, so that $W_e$ is included in $G_g$ if and only if there is an isomorphism between $\C$ and $\C/\langle g^{N/e}\rangle$ mapping the symmetry $g$ to $g'$.

One has the inclusions
\be \Gamma_0(N)\times \Gamma_0(N) \subseteq \Aut(L^{(\delta)})\subseteq \hat\Gamma_0(N)\times \hat\Gamma_0(N)\ ,
\ee where $\hat\Gamma_0(N)$ is the normalizer of $\Gamma_0(N)$ in $SL(2,\RR)$.

\vspace{.3cm}

\subsubsection{Self-dualities}

\noindent Consider now the subgroup $G_g \subseteq \Aut(L^{(\delta)})$ of automorphisms of $L^{(\delta)}$ such that
\be \C_{\gamma(r,s)}\cong \C_{r,s}\ .
\ee 
Equivalently, $G_g$ is the subgroup of elements in $\Aut(L^{(\delta)})$ for which the CFT $\C'$ is equivalent to $\C$ and $g=g'$. By definition, this is the group of self-dualities of the CHL model. 

Note that inside $G_g$ we also have $\Aut^0(L^{(\delta)})$, the subgroup of automorphisms that act trivially on $L^{(\delta)}/L^{(\delta)}_{0,0}$. These automorphisms map each $\C^{T^d}_{r,s}$ onto itself and they are all self-dualities of the CHL model. For $d=2$, it is easy to see by a direct calculation that
\begin{multline} \Aut^0(L^{(\delta)})=\left\{(\left(\begin{smallmatrix}
a & b\\
c & d
\end{smallmatrix}\right),\left(\begin{smallmatrix}
a' & b'\\
c' & d'
\end{smallmatrix}\right)) \in SL(2,\ZZ)\times SL(2,\ZZ) \mid c\equiv 0,\ c'\equiv 0,\ dd'\equiv 1\mod N \right\}\ .
\end{multline} Notice that we have the inclusions
\be\label{aut0delta} \Gamma_1(N)\times\Gamma_1(N) \subseteq \Aut^0(L^{(\delta)}) \subseteq \Gamma_0(N)\times\Gamma_0(N)\ .
\ee

\vspace{.3cm}

\noindent Finally, one has the obvious (normal) inclusions among the various automorphism groups:
\be \Aut^0(L^{(\delta)})\subseteq G_g\subseteq \Aut(L^{(\delta)})\ .
\ee

\subsubsection{Beyond T-duality}
The parent theory $\mathcal{S}[\mathcal{C}\times T^d]$, which is the starting point of the CHL-construction, always has a duality group containing $O(d,d,\ZZ)$ because of the $T^d$-factor. While we restrict the orbifold  to symmetries $g$ that commute with $O(d,d,\ZZ)$, it  is clear that the shift $\delta$ does \emph{not} commute with it. This implies that by conjugating $\delta$ by an $O(d,d,\ZZ)$ transformation, we obtain a different (but equivalent) CHL model, with a different (but isomorphic) $L^{(\delta)}$ lattice.

 To understand the full action of this group, let us first notice that $\delta$ can be defined as a null vector in $\frac{1}{N}L$; the corresponding shift is just the symmetry that multiplies a state with charges $X\in L$ by $e^{2\pi i (\delta,X)}$.  The symmetry defined by $\delta$ acts trivially if and only if $\delta\in L\subseteq \frac{1}{N}L$, so the CHL model only depends on the class $[\delta]$ in the quotient $\frac{1}{N}L/L\cong (\ZZ_N)^d$. Notice that the quadratic form on $L$ induces a quadratic form $q:\frac{1}{N}L/L\to \frac{1}{N}\ZZ/\ZZ$ defined by $q([\delta])=\frac{N}{2}(\delta,\delta)$. The symmetries we are interested in correspond to null vectors $\delta \in \frac{1}{N}L^\vee$, $(\delta,\delta)=0$ (otherwise the orbifold by $(\delta,g)$ does not satisfy the level matching condition), so that
\be\label{nullo} q([\delta])=0 \mod \ZZ\ .
\ee

Vice versa, if this condition is satisfied for a certain $[\delta]$, then there is always a representative $\delta \in \frac{1}{N}L$ which is null $(\delta,\delta)=0$. We can also restrict ourselves to the case where $[\delta]$ is exactly of order $N$, i.e. when $N\delta \in L$ is a primitive null vector, for any representative $\delta$ of $[\delta]$. Now, $O(d,d,\ZZ)$ (and also $SO^+(d,d,\ZZ)$) acts transitively on the set of primitive null vectors in $L$. Therefore, the classes $[\delta]$ form a single orbit under the induced action of  $O(d,d,\ZZ)$ on $\frac{1}{N}L/L$.  Given $[\delta]$, the untwisted sector lattice $L^{(\delta)}_{0,0}\subseteq L$ is defined as the sublattice `orthogonal to $[\delta]$', i.e.  the lattice of vectors $v\in L$ such that $(\delta,v)\in \ZZ$
\be L^{(\delta)}_{0,0}=\{v\in L\mid (v,\delta)\in \ZZ\}\ .
\ee It is clear that its dual $L^{(\delta)}$ contains $\delta$; in fact,  $L^{(\delta)}$ is generated by $L$ and $\delta$. The subgroup of $O(d,d,\ZZ)$ that leaves the class $[\delta]$ fixed is exactly the group $\Aut^0(L^{(\delta)})$ of automorphisms that act trivially on $L^{(\delta)}/L_{0,0}^{(\delta)}$. Therefore, the inequivalent classes $[\delta]\in \frac{1}{N}L/L$ are in one-to-one correspondence with the cosets
\be \{[\delta]\in \frac{1}{N}L/L\mid ord([\delta])=N,\ q([\delta])=0\}\leftrightarrow O(d,d,\ZZ)/\Aut^0(L^{(\delta)})\ .
\ee

%\subsection{Non-trivial multiplier}

\section{Atkin-Lehner dualities and $N$-modularity}
\label{sec:TS}
\noindent We shall now consider the case of four-dimensional $\mathcal{N}=4$ CHL-models in a little more detail. This concerns the chain of 
models mentioned in Examples 1 and 2 of section \ref{sec:CHLdef}, namely those were the parent theories are heterotic string theory on $T^4\times T^2$ or type IIA/B string theory on $K3\times T^2$. We shall in particular analyze $S$- and $T$-dualities of these models and discuss their consequences for the structure of the electric-magnetic charge lattices. 

\subsection{$T$- and $U$-dualities}
Consider type IIA string theory on $K3\times T^2$. This is a four-dimensional $\mathcal{N}=4$ string theory with duality group $SL(2,\mathbb{Z})\times O(6,22;\mathbb{Z})$, where the first factor is the $S$-duality group and the second factor is the $T$-duality group. The full moduli space of this theory is then:
\begin{align}
\mathcal{M}=SL(2,\mathbb{Z})\backslash SL(2,\mathbb{R})/SO(2)\times O(6,22;\mathbb{Z})\backslash O(6,22;\mathbb{R})/(O(6)\times O(22))
=\Gamma\backslash \T\ ,
\end{align} where the `Teichm\"uller space' $\T$ is the Grassmannian $SL(2,\mathbb{R})/SO(2)\times O(6,22;\mathbb{R})/(O(6)\times O(22))$ and $\Gamma$ is the suality group $SL(2,\ZZ)\times O(6,22,\ZZ)$.
At generic points in this moduli space the theory has a $U(1)^{28}$ gauge group and an associated electric-magnetic charge lattice 
\begin{equation}
L=\Gamma^{6,22}\oplus\Gamma^{6,22}=L_e\oplus L_m.
\end{equation}

\noindent We now take the orbifold of this $\mathcal{N}=4$ theory by a pair $(g,\delta)$,  such that: 
\begin{itemize}
\item $g$ is an order $N$ automorphism of the  $\C^{K3}$ which exists at points with generic gauge group $U(1)^{28}$ in $\mathcal{M}$, satisfying the level-matching condition $L_0-\bar L_0\in \frac{1}{N}\ZZ$ and commuting with the $\mathcal{N}=(4,4)$ algebra; 
\item $\delta$ is a translation by $1/N$ along the second circle of the torus factorization $T^2=S^1\times \tilde{S}^1$.
\end{itemize}

\noindent  The resulting CHL-model $\mathbb{CHL}_{g,\delta}(\text{IIA}, \C^{K3}, T^2)$ has a  moduli space $\mathcal{M}_{\mathbb{CHL}}$ which is again the quotient of a simply connected space $\T_{g,\delta}\subseteq \T$ by a discrete duality group $\Gamma_{g,\delta}$. The space $\T_{g,\delta}$ is a symmetric space of the form
\be \T_{g,\delta}=SL(2,\mathbb{R})/SO(2)\times O(6,d-6;\mathbb{R})/(O(6)\times O(d-6))\ ,
\ee where $8\le d\le 28$ is the dimension of the $g$-fixed subspace in $\Gamma^{6,22}\otimes \RR$.
 In particular, $T_{g,\delta}$ contains
\begin{equation}
( SL(2,\mathbb{R})/SO(2))_{S_{het}}\times   ( SL(2,\mathbb{R})/SO(2))_{T_{het}}\times  ( SL(2,\mathbb{R})/SO(2))_{U_{het}}\times \T^{K3}_{g}\subset \mathcal{T}_{\mathbb{CHL}},
\label{modspace}
\end{equation}
where the first $ SL(2,\mathbb{R})/SO(2)$ factor is parametrized by the heterotic axio-dilaton $S_{het}$, the second by the K\"ahler modulus $T_{het}$ of $T^2$  and the third factor by the complex structure modulus $U_{het}$. Furthermore, $\T^{K3}_g$ is the $g$-invariant part of $O(4,20;\mathbb{R})/(O(4)\times O(20))$ and parametrizes the K3 sigma models with a symmetry $g$. The full duality group $\Gamma_{g,\delta}$ is quite complicated: it obviously contains the subgroup of $SL(2,\ZZ)\times O(6,22,\ZZ)$ that leaves the pair $(\delta,g)$ invariant, but it is, in general, larger than that. In this article, we will use the techniques described in section \ref{s:CHLTdual} to derive a large group of duality. However, we are not able to determine whether this is the complete duality group of the CHL model -- to the best of our knowledge, this is still an open problem.

One apparent difficulty in this program is that in section \ref{s:CHLTdual} we only consider T-dualities, while now we are interested in determining the full group of (in general, non-perturbative) U-dualities.  Here, string-string duality comes to a help. Indeed, the same CHL model can be described in three equivalent ways as a heterotic, type IIA or type IIB compactification and what is consider a T-duality in one of these frames might be non-perturbative in the other ones. Thus, by combining the T-duality groups in the three different frames, we generate a large non-perturbative U-duality group.  
%In each of these duality frames, two of the $SL(2,\RR)/SO(2)$ factors in \eqref{modspace} combine into the $SO^+(2,2,\RR)$ moduli space parametrising geometry and B-field on $T^2$, while the third $SL(2,\RR)/SO(2)$ parametrizes the complexified string coupling constant. Therefore, we can perform the analysis of the (perturbative) T-duality group in each of these two pictures, following the general startegy outlined in section \ref{sec:TS}, and then combine them to obtain a non-perturbative U-duality group of the full theory.

As a starting point, we notice that the duality group contains a subgroup
\be \Gamma_1(N)_{S_{het}}\times \Gamma_1(N)_{T_{het}}\times \Gamma_1(N)_{U_{het}}\times C_{O(4,20,\ZZ)}(g)\ , 
\ee with each factor acting independently on the corresponding factor in \eqref{modspace}, where 
\begin{equation}
\Gamma_1(N)=\left\{\begin{pmatrix} a & b \\ c & d \end{pmatrix}\in SL(2,\mathbb{Z})\mid a\equiv 1\, \text{mod}\, N, \, c\equiv 0\, \text{mod}\, N\right\}.
\end{equation}
 and $C_{O(4,20,\ZZ)}(g)$ is the centralizer of $g$ in $O(4,20,\ZZ)$. This group can be easily extended to the group 
 \be \langle   \Gamma_1(N)^{\times 3}, C_{O(5,21,\ZZ)}(g)\rangle\ ,
 \ee 
 generated by the $\Gamma_1(N)^{\times 3}$ factor as well as the centralizer $C_{O(5,21,\ZZ)}(g)$ of $g$ in the larger group $O(5,21,\ZZ)$ that leaves invariant only the $\Gamma^{1,1}$ sublattice of winding-momenta associated with the circle  $S^1$ along the shift $\delta$, but possibly mixes the winding-momenta along the second circle of $T^2$ with other charges. Note that $C_{O(5,21,\ZZ)}(g)$ commutes with $\Gamma_1(N)_{S_{het}}$ but not with $\Gamma_1(N)_{T_{het}}\times \Gamma_1(N)_{U_{het}}$, so this is not a direct product of groups.

To make contact with the notation of section \ref{s:CHLTdual}, observe that each product $\Gamma_1(N)\times \Gamma_1(N)$ of a pair of factors can be interpreted as a subgroup of $\Aut^0(L_{(\delta)})$ (see eq.\eqref{aut0delta}) in each of the three duality frames. This suggests that the product $\Gamma_1(N)^{\times 3}$ can be extended to a subgroup of $\Gamma_0(N)^{\times 3}$, similarly to the way $\Gamma_1(N)\times\Gamma_1(N)$ is extended to $\Aut^0(L_{(\delta)})$. By combining the groups $\Aut^0(L_{(\delta)})$ in the three equivalent frames, one finds that the U-duality group must contain
 \be \langle (\Gamma_0(N)_{S_{het}}\times \Gamma_0(N)_{T_{het}}\times \Gamma_0(N)_{U_{het}})_0, C_{O(5,21,\ZZ)}(g)\rangle\ , 
\ee where 
\be  (\Gamma_0(N)^{\times 3})_0:=\{\left(\left(\begin{smallmatrix} a & b\\ c & d\end{smallmatrix}\right),\left(\begin{smallmatrix} a' & b'\\ c' & d'\end{smallmatrix}\right),\left(\begin{smallmatrix} a'' & b''\\ c'' & d''\end{smallmatrix}\right)\right)\in \Gamma_0(N)^{\times 3}\mid dd'd''\equiv 1\mod N \} \ .
\ee
%a discrete subgroup of $SL(2,\RR)^{\times 3}$ (The direct product of three copies of $SL(2,\RR)$, each acting on the corresponding factor). In the unorbifolded cases, the duality group is simply the product $SL(2,\ZZ)^{\times 3}$. Naively, in the orbifold theory we would expect this group to be broken  to the product of three copies of
%\begin{equation}
%\Gamma_1(N)=\left\{\begin{pmatrix} a & b \\ c & d \end{pmatrix}\in SL(2,\mathbb{Z})\mid a\equiv 1\, \text{mod}\, N, \, c\equiv 0\, \text{mod}\, N\right\}.
%\end{equation}
% What is surprising, however, is that $\Gamma_{g,\delta}$ can actually be larger and in general is not contained in $SL(2,\mathbb{Z})^{\times 3}$ and it is not a product of three groups acting independently on the three moduli.

Finally, one needs to extend $\Aut^0(L_{(\delta)})$ to the group of self-dualities $G_g$. We do this in two steps, by first extending to $\Gamma_0(N)^{\times 3}$ and then including the Atkin-Lehner involutions. In the first step, we notice that $\Gamma_0(N)/\Gamma_1(N)\cong (\ZZ/N\ZZ)^\times$ (the multiplicative group of elements in $\ZZ/N\ZZ$ that are coprime to $N$), with the isomorphism explicitly given by $\left(\begin{smallmatrix} a & b\\ c & d\end{smallmatrix}\right)\mapsto d$. Furthermore, if
\be \N(g)\equiv \N_{O(5,21,\ZZ)}(g):=\{h\in O(5,21,\ZZ)\mid \langle hgh^{-1}\rangle = \langle g\rangle\}\ ,
\ee is the normalizer of the cyclic group $\langle g\rangle$ in $O(5,21,\ZZ)$, then there is a homomorphism 
\be  a:\N(g)\to (\ZZ/N\ZZ)^\times\ ,
\ee with kernel $C_{O(5,21,\ZZ)}(g)$ defined by $ \N(g)\ni h\mapsto a(h)\in (\ZZ/N\ZZ)^\times$ if $hgh^{-1}=g^{a(h)}$. Combining these results, we notice that there is a homomorphism 
\be \phi:\langle \Gamma_0(N)^{\times 3}, \N(g)\rangle \to (\ZZ/N\ZZ)^\times\ ,
\ee that restricts to the homomorphisms above for the subgroups $\Gamma_0(N)^{\times 3}$ and $\N(g)$. The dualities $h\in \langle \Gamma_0(N)^{\times 3}, \N(g)\rangle$ map the group $(\delta,g)$ to $(\delta, g^{\phi(h)}\rangle$, so the kernel of the automorphism $\phi$ corresponds to self-dualities. Thus the U-duality group contains
\begin{align}\label{TdualPartial}
 &\langle \Gamma_0(N)^{\times 3}, \N(g)\rangle_0:=\ker\phi=\\
&\{\left(\left(\begin{smallmatrix} a & b\\ c & d\end{smallmatrix}\right),\left(\begin{smallmatrix} a' & b'\\ c' & d'\end{smallmatrix}\right),\left(\begin{smallmatrix} a'' & b''\\ c'' & d''\end{smallmatrix}\right),h\right)\in \Gamma_0(N)^{\times 3}\times \N(g)\mid dd'd''a(h)\equiv 1\mod N \}\ .
\end{align} Notice that in \cite{Persson:2015jka} it has been proved that the homomorphism $a:\N_{O(5,21,\ZZ)}(g)\to (\ZZ/N\ZZ)^\times$ is surjective, so that the projection $\langle \Gamma_0(N)^{\times 3}, \N(g)\rangle_0\to \Gamma_0(N)^{\times 3}$ is surjective as well. 

So far we only considered dualities that descend from the group $SL(2,\ZZ)\times O(6,22,\ZZ)$ of the parent theory. In general, this group can be enlarged by including the Atkin-Lehner involutions in the heterotic and type II pictures.  Let us choose a frame, for example the heterotic string frame, and consider the Atkin-Lenher T-dualities in this frame. The latter act only on the heterotic $T_{het}$ and $U_{het}$ moduli (the second and third factor in \eqref{modspace}) while leaving $S_{het}$ fixed and are generated by elements of the form 
\be (1,W^T_e,W^U_e)_0\in \hat\Gamma_0(N)_{S_{het}}\times \hat\Gamma_0(N)_{T_{het}}\times \hat\Gamma_0(N)_{U_{het}} ,\ee for $e$ an exact divisor of $N$. Here $W^T_e$ and $W^U_e$ are (possibly different) representatives in $\hat\Gamma_0(N)$ for the two Atkin-Lehner involutions relative to the exact divisor $e$, and the subscript $0$ means that the representatives are chosen in such a way that $W_e^TW_e^U\in \Gamma_1(N)$ rather than $\Gamma_0(N)$ (if this condition is not satisfied, than one needs to compose with some suitable $h\in \N(g)$, compatibly with \eqref{TdualPartial}).   The involution corresponding to the exact divisor $e$ should  be included in the U-duality group (i.e., it is a self-duality) if and only if the `internal' CFT $\C$ describing a heterotic string compactified on $T^4$  is isomorphic to the orbifold of $\C$ by $g^{N/e}$. The subset of exact divisors $e$ for which $W_e$ is a heterotic self-duality will be denoted by $H_{het}$ and form a subgroup of the group $H(N)$ of exact divisors of order $N$.\footnote{We recall that the set $H(N)$ of exact divisors of $N\in \NN$ can be endowed with a natural structure of finite abelian group of exponent $2$, with composition law
\be e*f:=\frac{ef}{\gcd(e,f)^2}\ .
\ee } It was argued in \cite{Persson:2015jka} that in every CHL model, all $W_e$ are heterotic self-dualities, so that $H_{het}=H(N)$.  Similarly, Atkin-Lehner T-dualities in the type IIA frame act only on the $S_{het}$ and $U_{het}$ moduli, while leaving $T_{het}$ fixed and are of the form
\be (W^S_e,W^T_e,1)_0\in \hat\Gamma_0(N)_{S_{het}}\times \hat\Gamma_0(N)_{T_{het}}\times \hat\Gamma_0(N)_{U_{het}} . \ee This transformation should be included in the U-duality group if and only if the internal CFT $\C$, which is a type II non-linear sigma model on K3, is isomorphic to the orbifold of $\C$ by $g^{N/e}$. Finally, a similar Atkin-Lehner T-dualities in type IIB have the form
\be (W^S_e,1,W^U_e)_0\in \hat\Gamma_0(N)_{S_{het}}\times \hat\Gamma_0(N)_{T_{het}}\times \hat\Gamma_0(N)_{U_{het}} , \ee and the set of exact divisors for which these transformation belongs to the U-duality group is the same as for type IIA. We denote by $H_{II}\subset H(N)$ the group of self-dualities in this case. Such groups depend on the particular CHL model and were discussed in \cite{Persson:2015jka} (See section \ref{s:Witte}).

By combining the analysis for the heterotic and type II descriptions, we conclude that the U-duality contains generators of the form
\be (W^S_e,W^T_{e'},W^U_{e''})_0\in \hat\Gamma_0(N)_{S_{het}}\times \hat\Gamma_0(N)_{T_{het}}\times \hat\Gamma_0(N)_{U_{het}} , \ee
with
\be (e,e',e'')\in H_{II}\times H(N)\times H(N) ,\qquad e*e'*e''=1\ .
\ee 
We conclude that the U-duality group contains a subgroup of the form
\begin{multline}\label{TdualFinal} \left\{\left(\left(\begin{smallmatrix} a\sqrt{e} & b/\sqrt{e}\\ cN/\sqrt{e} & d\sqrt{e}\end{smallmatrix}\right),\left(\begin{smallmatrix} a'\sqrt{e'} & b'/\sqrt{e'}\\ c'N'/\sqrt{e'} & d'\sqrt{e'}\end{smallmatrix}\right),\left(\begin{smallmatrix} a''\sqrt{e''} & b''/\sqrt{e''}\\ c''N''/\sqrt{e''} & d''\sqrt{e''}\end{smallmatrix}\right),h\right)\in \hat\Gamma_0(N)^{\times 3}\times \N(g)\mid \right.\\\left.dd'd''a(h)\equiv 1\mod N,\ (e,e',e'')\in H_{II}\times H(N)\times H(N),\ e*e'*e''=1 \right\} 
\end{multline}

\subsection{Witten index}\label{s:Witte}

In order to specify the subgroup $H_{II}$ in \eqref{TdualFinal}, one needs to determine under which conditions the orbifold $\mathcal{C}^{\prime}=\C^{K3}/\left<g\right>$ is a K3-model. This can be done by evaluating the Witten index 
\begin{equation}
I_{\mathcal{C}^{\prime}}=\text{Tr}_{\mathcal{C}^{\prime},  RR}(-1)^{F_L+F_R}.
\end{equation}
Recall that an order $N$ symmetry $g\in Co_0$ can be characterized by its \emph{Frame shape}, which encodes its eigenvalues in the defining 24-dimensional representation in $O(\Gamma^{4,20})$:
\begin{equation}
\prod_{a|N}a^{m(a)},
\end{equation}
where $m(a)\in \mathbb{Z}$ satisfies
\begin{equation}
\sum_{a|N} m(a) a=24.
\end{equation}
One can then show that the Witten index of $\mathcal{C}^{\prime}=\mathcal{C}^{K3}/\left<g\right>$ is given by \cite{Persson:2015jka}
\begin{equation}
I_{\mathcal{C}^{\prime}}=\sum_{a|N} m(N/a)a.
\end{equation}
From this one can conclude that there are three possibilities for the orbifold $\mathcal{C}^{\prime}$:
\begin{itemize}
\item $\mathcal{C}^{\prime}$ is a superconformal field theory on K3, 
\begin{equation}
I_{\mathcal{C}^{\prime}}=24,
\end{equation}
and the two symmetries $(g, \mathcal{Q})$ have the \emph{same} Frame shape;
\item
$\mathcal{C}^{\prime}$ is a superconformal field theory on K3, 
\begin{equation}
I_{\mathcal{C}^{\prime}}=24,
\end{equation}
and the two symmetries $(g, \mathcal{Q})$ have \emph{different} Frame shapes; 
\item $\mathcal{C}^{\prime}$ is a superconformal field theory on $T^4$, 
\begin{equation}
I_{\mathcal{C}^{\prime}}=0.
\end{equation}
\end{itemize}

Note that the frame shape of the quantum symmetry $\mathcal{Q}$ is given by $\prod_{a|N}a^{m(N/a)}$ and hence $(g,\mathcal{Q})$ can only have the same frame shape if the following condition holds
\begin{equation}
m(a)=m(N/a),
\end{equation}
in which case one says that $g$ has \emph{balanced} frame shape. If this holds, we say that the CHL-model is \emph{self-dual}, since then $\mathbb{CHL}_{\mathcal{Q},\delta^{\prime}}(\text{IIA}, \C^{K3}/\left<g\right>, {T^{\prime}}^2)$ and $\mathbb{CHL}_{g,\delta}(\text{IIA}, \C^{K3}, T^2)$ are in the same connected component of the moduli space $\mathcal{M}$. In contrast, in the second case, when $(g,\mathcal{Q})$ have different frame shapes, Fricke T-duality relates two \emph{inequivalent} K3 CHL-models.
The third case is very different from the others since the image of a K3 CHL-model is now a model based on an orbifold of $\mathcal{C}^{T^4}\times \mathcal{C}^{T^2}$.

\subsection{$N$-modularity}
The conclusion of the previous sections is that $\mathcal{N}=4$ CHL-models have duality groups which are larger than what is naively expected. In particular, the S- and T-duality groups contain the Atkin-Lehner involutions $W_e$, which belong to $SL(2,\mathbb{R})$ but are not contained in the $SL(2,\mathbb{Z})$-symmetry of the parent theory. In this section we will show that this duality symmetry yield strong constraints on the lattice of electric-magnetic charges.

Any $\mathcal{N}=4$ CHL-model of the form $\mathbb{CHL}_{{g},\delta}(\mathcal{S}, \C^{X}, {T}^2)$, with $X$ either $T^4$ or $K3$, has a lattice of electric-magnetic charges
\begin{equation}
L^{(g,\delta)}=L^{(g,\delta)}_e\oplus L^{(g,\delta)}_{m}\subset \Gamma^{6,22}\oplus \Gamma^{6,22}.
\end{equation}
The S-duality group $\Gamma_g$  acts on any vector $(Q, P)\in L^{(g,\delta)}$ by
\begin{equation}
\begin{pmatrix} Q \\ P \end{pmatrix} \longmapsto \begin{pmatrix} Q^{\prime} \\ P^{\prime} \end{pmatrix} =\begin{pmatrix} d & -b \\ -c & a \\ \end{pmatrix} \begin{pmatrix} Q \\ P \end{pmatrix},\qquad \quad \begin{pmatrix} a & b \\ c & d \\ \end{pmatrix} \in \Gamma_g,
\end{equation}
while at the same time acting on the axio-dilaton $S$ in the standard fractional way
\begin{equation}
S\longmapsto S^{\prime}=\frac{aS+b}{cS+d}.
\end{equation}
If $\mathbb{CHL}_{{g},\delta}(\mathcal{S}, \C^{X}, {T}^2)$ is self-dual (the frame shape of $g$ is balanced) then this is a symmetry of the theory. Consider in particular the action of the Fricke involution $W_N$:
\begin{equation}
S\longmapsto -\frac{1}{NS}, \qquad \qquad \begin{pmatrix} Q \\ P \end{pmatrix} \longmapsto  \begin{pmatrix} Q^{\prime} \\ P^{\prime} \end{pmatrix} =\begin{pmatrix} \tfrac{1}{\sqrt{N}} P \\ -\sqrt{N}Q \end{pmatrix}.
\label{Frickeduality}
\end{equation}
The charges $(Q^{\prime}, P^{\prime})$ belong to the Fricke dual lattice 
\begin{equation}
W_N\, :\, L^{(g, \delta)}\, \longrightarrow \, {L^{\prime}}^{(g, \delta)}={L^{\prime}}_e^{(g, \delta)}\oplus {L^{\prime}}_m^{(g, \delta)}.
\end{equation}
The Fricke S-duality \eqref{Frickeduality} then implies that the lattices are related as follows:
\begin{equation}
{L^{\prime}}_e^{(g, \delta)}={L}_m^{(g, \delta)}(1/N), \qquad \qquad {L^{\prime}}_m^{(g, \delta)}={L}_e^{(g, \delta)}(N).
\label{Frickelattices}
\end{equation}
Here, the notation $L(n)$ means that each vector in $L$ is rescaled by a factor $\sqrt{n}$ such that its quadratic form is rescaled by $n$. We further know that by standard electric-magnetic duality we must have 
\begin{equation}
{L}_e^{(g, \delta)}\cong {L^{*}}_m^{(g, \delta)},
\label{emduality}
\end{equation}
where $L^{*}$ denotes the standard dual lattice to $L$. If we consider a CHL-model which is self-dual under Fricke S-duality then we must further have
\begin{equation}
{L^{\prime}}_e^{(g, \delta)}\cong {L}_e^{(g, \delta)}, \qquad \qquad {L^{\prime}}_m^{(g, \delta)}\cong {L}_m^{(g, \delta)}.
\end{equation}
Combining this with the relation \eqref{emduality}, we deduce that Fricke S-duality \eqref{Frickelattices} enforces the following constraint on the electric and magnetic charge lattices:
\begin{equation}
{L}_e^{(g, \delta)}={L^{*}}_e^{(g, \delta)}(1/N), \qquad \qquad {L}_m^{(g, \delta)}={L^{*}}_m^{(g, \delta)}(N),
\end{equation}
namely that they should be isometric to their duals up to a rescaling of the quadratic form. In general, lattices that satisfy $L\cong L^{*}(N)$ are known as $N${\it-modular}. Beyond the dimension 2 case, $N$-modular lattices are very rare and the fact that the electric-magnetic charge lattices of CHL-models are required to be $N$-modular is a very strong prediction of Fricke, or more generally, Atkin-Lehner S-duality. 

Let us take a closer look at the implications of this $N$-modularity. We can write the lattices of $\mathbb{CHL}_{{g},\delta}(\mathcal{S}, \C^{X}, {T}^2)$ as 
\begin{eqnarray}
{L}_e^{(g, \delta)}&=& \begin{pmatrix}  & 1 \\ 1 &  \\ \end{pmatrix} \oplus \begin{pmatrix}  & \tfrac{1}{N} \\ \tfrac{1}{N} &  \\ \end{pmatrix} \oplus \left[(\Gamma^{4,20})^g\right]^{*}
\nonumber \\
{L}_m^{(g, \delta)}&=& \begin{pmatrix}  & 1 \\ 1 &  \\ \end{pmatrix} \oplus \begin{pmatrix}  & N \\ N &  \\ \end{pmatrix} \oplus (\Gamma^{4,20})^g,
\end{eqnarray}
where the first two factors represent the quadratic forms of the lattices associated with $T^2=S^1\times \tilde{S}^1$. Fricke S-duality now implies the following non-trivial $N$-modularity condition
\begin{equation}
(\Gamma^{4,20})^g\cong \sqrt{N}\left[(\Gamma^{4,20})^g\right]^{*}.
\label{nontrivialNarain}
\end{equation}
To give an example of what this constraint entails, consider $\mathbb{CHL}_{{g},\delta}(\text{IIA}, \C^{K3}, {T}^2)$ and choose $g$ to be the Conway element with frame shape $1^8 2^8$. This is the so-called \emph{Nikulin involution} which is an order 2 symplectic automorphism of K3. On the heterotic side this is the involution which exchanges the two $E_8$-factors. The electric-magnetic lattices take the form
\begin{eqnarray}
{L}_e^{(1^8 2^8, \delta)}&=& \begin{pmatrix}  & 1 \\ 1 &  \\ \end{pmatrix} \oplus \begin{pmatrix}  & \tfrac{1}{2} \\ \tfrac{1}{2} &  \\ \end{pmatrix} \oplus \Gamma^{4,4}\oplus E_8(-1/2)
\nonumber \\
{L}_m^{(1^8 2^8, \delta)}&=& \begin{pmatrix}  & 1 \\ 1 &  \\ \end{pmatrix} \oplus \begin{pmatrix}  & 2 \\ 2 &  \\ \end{pmatrix} \oplus \Gamma^{4,4}\oplus E_8(-2).
\end{eqnarray}
Fricke S-duality implies that there must be an isomorphism
\begin{equation}
\Gamma^{4,4}\oplus E_8(-2)\cong \Gamma^{4,4}(2)\oplus E_8(-1).
\end{equation}
One can prove that this indeed holds, quite non-trivially. In fact, in \cite{Persson:2015jka} we verified that \eqref{nontrivialNarain} is satisfied for all self-dual CHL-models, i.e. for those $g$ whose frame shapes are balanced. 

More generally, if for a CHL model $\mathbb{CHL}_{{g},\delta}$, with $g$ of order $N$, the group $H_{II}$ contains an exact divisor $e$ of $N$, then there is a duality acting on the heterotic $S$-modulus and on the electric and magnetic charges by
\be S\longmapsto \frac{aeS+b}{NcS+de}, \qquad \qquad \begin{pmatrix} Q \\ P \end{pmatrix} \longmapsto  \begin{pmatrix} Q^{\prime} \\ P^{\prime} \end{pmatrix} =\begin{pmatrix} a\sqrt{e}Q+\tfrac{b}{\sqrt{e}} P \\ c\frac{N}{\sqrt{e}}Q+d\sqrt{e}P \end{pmatrix}.
\ee Therefore, we expect isomorphisms
\be\label{ALmodularity} L_e\cong {\rm span}_{\ZZ}( \sqrt{e}L_e \cup \frac{1}{\sqrt{e}}
L_m)\qquad L_m\cong {\rm span}_{\ZZ}( \frac{N}{\sqrt{e}}L_e \cup \sqrt{e}
L_m).
\ee Some of these isomorphisms have been verified in \cite{Persson:2015jka}.

We conclude by noting that \eqref{nontrivialNarain} has consequences for the integral cohomology lattice of K3-surfaces, as was pointed out in \cite{Persson:2015jka}. Consider the $g$-invariant subspace $H^{\text{even}}(K3; \mathbb{Z})^g$ of the even integral cohomology of a K3-surface. If the model is invariant under Fricke S-duality, then $H^{\text{even}}(K3; \mathbb{Z})^g$  must be $N$-modular:
\begin{equation}
H^{\text{even}}(K3; \mathbb{Z})^g=\sqrt{N}\left[H^{\text{even}}(K3; \mathbb{Z})^g\right]^{*}.
\end{equation} This is true, in particular, whenever $g$ is `geometric', i.e. it is induced by some symplectic automorphism of the target space K3. In this case, the CHL model is also invariant under all Atkin-Lehner dualities, so that $H^{\text{even}}(K3; \mathbb{Z})^g$ must satisfy also isomorphisms of the form \ref{ALmodularity}.
Such $g$-invariant sublattices of $H^{\text{even}}(K3;\mathbb{Z})$ were studied in \cite{GarbagnatiSarti2009} but the $N$-modularity appears to have gone unnoticed in the literature. 

\section{Discussion}
\label{sec_disc}

\noindent In this note we have given a general construction of CHL-models starting from any string theory compactified on some internal CFT $\mathcal{C}\times T^d$. This  provides a vast generalization of our previous work on $\mathcal{N}=4$ and monstrous CHL-models \cite{Persson:2015jka,Paquette:2016xoo,Paquette:2017xui}, showing that the features discovered therein, such as Fricke T-duality and $N$-modularity, occur universally. 

The monstrous CHL-models constructed in \cite{Paquette:2016xoo,Paquette:2017xui} provided new insight into monstrous moonshine, in particular by giving a novel physics derivation of its so called genus zero property. It would be interesting to analyze whether there are more examples of such CHL-models that  could be of similar use in understanding other types of moonshine phenomena. For example, one interesting case would be to take type II string theory on $V^{s\natural}\times \overline{V}^{s\natural}\times T^2$, where $V^{s\natural}$ is Duncan's super-module \cite{2005math......2267D} associated with moonshine for the Conway group $Co_0$. This might potentially provide a spacetime interpretation of the genus zero property of Conway moonshine\footnote{We thank Timm Wrase for suggesting this example, and for related discussions.}, as well as shed light on the associated algebra of BPS-states. 

Another interesting sector to analyze in more detail is type II string theory on Calabi-Yau 3-folds. This would give a class of  CHL-models that should  give new insight into the $\mathcal{N}=2$ Mathieu moonhine observed in \cite{2005math......2267D}. Such a construction might provide a connection between the Fricke symmetries of CHL-models and those observed in topological strings \cite{Alim:2013eja}. 

The Fricke S-dualities  in $\mathcal{N}=4$ CHL-models act as $S\to -1/(NS)$ on the heterotic axio-dilaton $S$, where $N$  is the order of the orbifold symmetry. This is reminiscent of the  duality $\tau \to -1/(m\tau)$ occurring in the gauge theory approach to the geometric Langlands program \cite{Kapustin:2006pk}. In this case $\tau$ is the complex gauge coupling and $m$ depends om the gauge group $G$. Recently \cite{Aganagic:2017smx}, this was interpreted as a duality in $(2,0)$  little string theory with defects, in which case $m$  corresponds to the order of a subgroup $H\subset G$ which is responsible for a twist around the complex plane that supports the defect. In our $\mathcal{N}=4$ CHL-models the gauge group is generically $U(1)^{28}$ but is enhanced to a non-abelian group at certain singular loci in the moduli space. It would be interesting to understand whether there is any relation to the Fricke S-dualities in CHL-models at these singular points and geometric Langlands duality of $(2,0)$-theories. 

%\begin{itemize}
%\item Similar stuff should work starting from $L$ of signature $(2,3)$. The group of dualities will be the modular group of $\Phi_{e,g}$...
%\end{itemize}

\section*{Acknowledgements}
\noindent We are grateful to Terry Gannon, Natalie Paquette, Boris Pioline and Timm Wrase for discussions and correspondence. We also thank the organizers of the Durham Symposium on ``New Moonshine, Mock Modular Forms and String Theory'', where this work was presented. We also thank the Simons Center for Geometry and Physics for hospitality while this work was being finalized. RV is supported by a grant from Programma per Giovani Ricercatori `Rita Levi Montalcini', and thanks SLAC and the Stanford Institute for Theoretical Physics for hospitality.

\bibliographystyle{utphys}

\bibliography{Refs}

\providecommand{\href}[2]{#2}\begingroup\raggedright\begin{thebibliography}{10}

\bibitem{Shih:2005uc}
D.~Shih, A.~Strominger, and X.~Yin, ``{Recounting Dyons in N=4 string
  theory},'' \href{http://dx.doi.org/10.1088/1126-6708/2006/10/087}{{\em JHEP}
  {\bfseries 10} (2006) 087},
\href{http://arxiv.org/abs/hep-th/0505094}{{\ttfamily arXiv:hep-th/0505094
  [hep-th]}}.
%%CITATION = HEP-TH/0505094;%%.

\bibitem{Dabholkar:2011ec}
A.~Dabholkar, J.~Gomes, and S.~Murthy, ``{Localization \& Exact Holography},''
  \href{http://dx.doi.org/10.1007/JHEP04(2013)062}{{\em JHEP} {\bfseries 04}
  (2013) 062},
\href{http://arxiv.org/abs/1111.1161}{{\ttfamily arXiv:1111.1161 [hep-th]}}.
%%CITATION = ARXIV:1111.1161;%%.

\bibitem{Dabholkar:2010uh}
A.~Dabholkar, J.~Gomes, and S.~Murthy, ``{Quantum black holes, localization and
  the topological string},''
  \href{http://dx.doi.org/10.1007/JHEP06(2011)019}{{\em JHEP} {\bfseries 06}
  (2011) 019},
\href{http://arxiv.org/abs/1012.0265}{{\ttfamily arXiv:1012.0265 [hep-th]}}.
%%CITATION = ARXIV:1012.0265;%%.

\bibitem{Dabholkar:2014ema}
A.~Dabholkar, J.~Gomes, and S.~Murthy, ``{Nonperturbative black hole entropy
  and Kloosterman sums},''
  \href{http://dx.doi.org/10.1007/JHEP03(2015)074}{{\em JHEP} {\bfseries 03}
  (2015) 074},
\href{http://arxiv.org/abs/1404.0033}{{\ttfamily arXiv:1404.0033 [hep-th]}}.
%%CITATION = ARXIV:1404.0033;%%.

\bibitem{Murthy:2015yfa}
S.~Murthy and V.~Reys, ``{Functional determinants, index theorems, and exact
  quantum black hole entropy},''
  \href{http://dx.doi.org/10.1007/JHEP12(2015)028}{{\em JHEP} {\bfseries 12}
  (2015) 028},
\href{http://arxiv.org/abs/1504.01400}{{\ttfamily arXiv:1504.01400 [hep-th]}}.
%%CITATION = ARXIV:1504.01400;%%.

\bibitem{Murthy:2015zzy}
S.~Murthy and V.~Reys, ``{Single-centered black hole microstate degeneracies
  from instantons in supergravity},''
  \href{http://dx.doi.org/10.1007/JHEP04(2016)052}{{\em JHEP} {\bfseries 04}
  (2016) 052},
\href{http://arxiv.org/abs/1512.01553}{{\ttfamily arXiv:1512.01553 [hep-th]}}.
%%CITATION = ARXIV:1512.01553;%%.

\bibitem{Dabholkar:2012nd}
A.~Dabholkar, S.~Murthy, and D.~Zagier, ``{Quantum Black Holes, Wall Crossing,
  and Mock Modular Forms},''
\href{http://arxiv.org/abs/1208.4074}{{\ttfamily arXiv:1208.4074 [hep-th]}}.
%%CITATION = ARXIV:1208.4074;%%.

\bibitem{Chaudhuri:1995fk}
S.~Chaudhuri, G.~Hockney, and J.~D. Lykken, ``{Maximally supersymmetric string
  theories in $D < 10$},''
  \href{http://dx.doi.org/10.1103/PhysRevLett.75.2264}{{\em Phys.Rev.Lett.}
  {\bfseries 75} (1995) 2264--2267},
\href{http://arxiv.org/abs/hep-th/9505054}{{\ttfamily arXiv:hep-th/9505054
  [hep-th]}}.
%%CITATION = HEP-TH/9505054;%%.

\bibitem{David:2006ji}
J.~R. David, D.~P. Jatkar, and A.~Sen, ``{Product representation of Dyon
  partition function in CHL models},''
  \href{http://dx.doi.org/10.1088/1126-6708/2006/06/064}{{\em JHEP} {\bfseries
  06} (2006) 064},
\href{http://arxiv.org/abs/hep-th/0602254}{{\ttfamily arXiv:hep-th/0602254
  [hep-th]}}.
%%CITATION = HEP-TH/0602254;%%.

\bibitem{Sen:2009md}
A.~Sen, ``{A Twist in the Dyon Partition Function},''
  \href{http://dx.doi.org/10.1007/JHEP05(2010)028}{{\em JHEP} {\bfseries 05}
  (2010) 028},
\href{http://arxiv.org/abs/0911.1563}{{\ttfamily arXiv:0911.1563 [hep-th]}}.
%%CITATION = ARXIV:0911.1563;%%.

\bibitem{Sen:2010ts}
A.~Sen, ``{Discrete Information from CHL Black Holes},''
  \href{http://dx.doi.org/10.1007/JHEP11(2010)138}{{\em JHEP} {\bfseries 1011}
  (2010) 138},
\href{http://arxiv.org/abs/1002.3857}{{\ttfamily arXiv:1002.3857 [hep-th]}}.
%%CITATION = ARXIV:1002.3857;%%.

\bibitem{Govindarajan:2010fu}
S.~Govindarajan, ``{BKM Lie superalgebras from counting twisted CHL dyons},''
  \href{http://dx.doi.org/10.1007/JHEP05(2011)089}{{\em JHEP} {\bfseries 1105}
  (2011) 089},
\href{http://arxiv.org/abs/1006.3472}{{\ttfamily arXiv:1006.3472 [hep-th]}}.
%%CITATION = ARXIV:1006.3472;%%.

\bibitem{Govindarajan:2011em}
S.~Govindarajan, ``{Unravelling Mathieu Moonshine},''
  \href{http://dx.doi.org/10.1016/j.nuclphysb.2012.07.005}{{\em Nucl.Phys.}
  {\bfseries B864} (2012) 823--839},
\href{http://arxiv.org/abs/1106.5715}{{\ttfamily arXiv:1106.5715 [hep-th]}}.
%%CITATION = ARXIV:1106.5715;%%.

\bibitem{Cheng:2010pq}
M.~C. Cheng, ``{K3 Surfaces, N=4 Dyons, and the Mathieu Group M24},''
  \href{http://dx.doi.org/10.4310/CNTP.2010.v4.n4.a2}{{\em
  Commun.Num.Theor.Phys.} {\bfseries 4} (2010) 623--658},
\href{http://arxiv.org/abs/1005.5415}{{\ttfamily arXiv:1005.5415 [hep-th]}}.
%%CITATION = ARXIV:1005.5415;%%.

\bibitem{Gaberdiel:2012gf}
M.~R. Gaberdiel, D.~Persson, H.~Ronellenfitsch, and R.~Volpato, ``{Generalized
  Mathieu Moonshine},''
  \href{http://dx.doi.org/10.4310/CNTP.2013.v7.n1.a5}{{\em Commun.Num.Theor
  Phys.} {\bfseries 07} (2013) 145--223},
\href{http://arxiv.org/abs/1211.7074}{{\ttfamily arXiv:1211.7074 [hep-th]}}.
%%CITATION = ARXIV:1211.7074;%%.

\bibitem{Gaberdiel:2013nya}
M.~R. Gaberdiel, D.~Persson, and R.~Volpato, ``{Generalised Moonshine and
  Holomorphic Orbifolds},''
\href{http://arxiv.org/abs/1302.5425}{{\ttfamily arXiv:1302.5425 [hep-th]}}.
%%CITATION = ARXIV:1302.5425;%%.

\bibitem{Persson:2013xpa}
D.~Persson and R.~Volpato, ``{Second Quantized Mathieu Moonshine},''
  \href{http://dx.doi.org/10.4310/CNTP.2014.v8.n3.a2}{{\em
  Commun.Num.TheorPhys.} {\bfseries 08} (2014) 403--509},
\href{http://arxiv.org/abs/1312.0622}{{\ttfamily arXiv:1312.0622 [hep-th]}}.
%%CITATION = ARXIV:1312.0622;%%.

\bibitem{Persson:2015jka}
D.~Persson and R.~Volpato, ``{Fricke S-duality in CHL models},''
  \href{http://dx.doi.org/10.1007/JHEP12(2015)156}{{\em JHEP} {\bfseries 12}
  (2015) 156},
\href{http://arxiv.org/abs/1504.07260}{{\ttfamily arXiv:1504.07260 [hep-th]}}.
%%CITATION = ARXIV:1504.07260;%%.

\bibitem{K3symm}
M.~R. Gaberdiel, S.~Hohenegger, and R.~Volpato, ``{Symmetries of K3 sigma
  models},'' \href{http://dx.doi.org/10.4310/CNTP.2012.v6.n1.a1}{{\em
  Commun.Num.Theor.Phys.} {\bfseries 6} (2012) 1--50},
\href{http://arxiv.org/abs/1106.4315}{{\ttfamily arXiv:1106.4315 [hep-th]}}.
%%CITATION = ARXIV:1106.4315;%%.

\bibitem{Bossard:2017wum}
G.~Bossard, C.~Cosnier-Horeau, and B.~Pioline, ``{Four-derivative couplings and
  BPS dyons in heterotic CHL orbifolds},''
\href{http://arxiv.org/abs/1702.01926}{{\ttfamily arXiv:1702.01926 [hep-th]}}.
%%CITATION = ARXIV:1702.01926;%%.

\bibitem{Paquette:2016xoo}
N.~M. Paquette, D.~Persson, and R.~Volpato, ``{Monstrous BPS-Algebras and the
  Superstring Origin of Moonshine},''
\href{http://arxiv.org/abs/1601.05412}{{\ttfamily arXiv:1601.05412 [hep-th]}}.
%%CITATION = ARXIV:1601.05412;%%.

\bibitem{FLM}
I.~Frenkel, J.~Lepowsky, and A.~Meurman, {\em Vertex operator algebras and the
  Monster}, vol.~134.
\newblock Academic press, 1989.

\bibitem{duncan2007super}
J.~F. Duncan, ``Super-moonshine for {Conway}'s largest sporadic group,'' {\em
  Duke Mathematical Journal} {\bfseries 139} no.~2, (2007) 255--315.

\bibitem{VafaWitten95}
C.~Vafa and E.~Witten, ``{Dual string pairs with N=1 and N=2 supersymmetry in
  four-dimensions},''
  \href{http://dx.doi.org/10.1016/0920-5632(96)00025-4}{{\em
  Nucl.Phys.Proc.Suppl.} {\bfseries 46} (1996) 225--247},
\href{http://arxiv.org/abs/hep-th/9507050}{{\ttfamily arXiv:hep-th/9507050
  [hep-th]}}.
%%CITATION = HEP-TH/9507050;%%.

\bibitem{GarbagnatiSarti2009}
A.~Garbagnati and A.~Sarti, ``Elliptic fibrations and symplectic automorphisms
  on {$K3$} surfaces,'' \href{http://dx.doi.org/10.1080/00927870902828785}{{\em
  Comm. Algebra} {\bfseries 37} no.~10, (2009) 3601--3631}.
  \url{http://dx.doi.org/10.1080/00927870902828785}.

\bibitem{Paquette:2017xui}
N.~M. Paquette, D.~Persson, and R.~Volpato, ``{BPS Algebras, Genus Zero, and
  the Heterotic Monster},''
\href{http://arxiv.org/abs/1701.05169}{{\ttfamily arXiv:1701.05169 [hep-th]}}.
%%CITATION = ARXIV:1701.05169;%%.

\bibitem{2005math......2267D}
J.~F. {Duncan}, ``{Super-moonshine for Conway's largest sporadic group},'' {\em
  ArXiv Mathematics e-prints} (Feb., 2005) ,
  \href{http://arxiv.org/abs/math/0502267}{{\ttfamily math/0502267}}.

\bibitem{Alim:2013eja}
M.~Alim, E.~Scheidegger, S.-T. Yau, and J.~Zhou, ``{Special Polynomial Rings,
  Quasi Modular Forms and Duality of Topological Strings},''
  \href{http://dx.doi.org/10.4310/ATMP.2014.v18.n2.a4}{{\em Adv. Theor. Math.
  Phys.} {\bfseries 18} no.~2, (2014) 401--467},
\href{http://arxiv.org/abs/1306.0002}{{\ttfamily arXiv:1306.0002 [hep-th]}}.
%%CITATION = ARXIV:1306.0002;%%.

\bibitem{Kapustin:2006pk}
A.~Kapustin and E.~Witten, ``{Electric-Magnetic Duality And The Geometric
  Langlands Program},''
  \href{http://dx.doi.org/10.4310/CNTP.2007.v1.n1.a1}{{\em Commun. Num. Theor.
  Phys.} {\bfseries 1} (2007) 1--236},
\href{http://arxiv.org/abs/hep-th/0604151}{{\ttfamily arXiv:hep-th/0604151
  [hep-th]}}.
%%CITATION = HEP-TH/0604151;%%.

\bibitem{Aganagic:2017smx}
M.~Aganagic, E.~Frenkel, and A.~Okounkov, ``{Quantum q-Langlands
  Correspondence},''
\href{http://arxiv.org/abs/1701.03146}{{\ttfamily arXiv:1701.03146 [hep-th]}}.
%%CITATION = ARXIV:1701.03146;%%.

\end{thebibliography}\endgroup

\end{document}